\documentclass[12pt]{article}

\usepackage[left=2cm,right=2cm,top=2cm,bottom=2cm]{geometry}

\usepackage{graphicx}
\usepackage{amsmath}
\usepackage{color}
\usepackage{cuted}

\usepackage[bookmarks=true,colorlinks=true,citecolor=blue,linkcolor=red,urlcolor=magenta]{hyperref}

\usepackage{soul}

\usepackage[numbers,sort&compress]{natbib}

\title{Neutron moderation spectrum considering inelastic scattering}
\author{V.D. Rusov$^a$\thanks{Corresponding author: siiis@te.net.ua}, V.A. Tarasov$^a$, S.A. Chernezhenko$^a$,\\
A.A. Kakaev$^a$, V.V. Urbanevich$^a$, V.M. Vashchenko$^b$}

\begin{document}

\date{
${}^a$\small{\emph{Odessa National Polytechnic University, Shevchenko av., 1, 65044, Odessa, Ukraine.}}
\\
${}^b$\small{\emph{State Ecological Academy of Post-Graduate Education and Management,\\ V. Lypkivskogo str., 35, bldg.2, 03035, Kyiv, Ukraine}}
\vspace{-5ex}
}

\maketitle

\abstract{
For the first time an analytic expression was obtained for the
inelastic neutron scattering law with an isotropic neutron source within the
gas model, considering moderating medium temperature as a parameter.
The inelastic scattering law is obtained, based on the solution of the
kinematic problem of neutron inelastic scattering on a nucleus in laboratory
coordinate system ($L$-system) in general case. I.e. in case not only a neutron
but also a nucleus have arbitrary velocity vector in $L$-system. 
Analytic expressions are found for the neutron flux density and moderation
spectrum in reactor fissile medium, both in case of the elastic scattering law,
obtained earlier by the authors, and in case of the inelastic scattering law
obtained in this paper. Both elastic and inelastic scattering laws are
considered to be dependent on the medium temperature. The obtained expressions
for neutron moderation spectra enable reinterpretation of physical nature of
the processes that determine the shape of neutron spectrum in a wide energy range.
%
%Keywords:~\emph{nuclear, nucleus, neutron, moderation, inelastic, elastic, scattering, temperature, kinematics, spectrum, interaction, cross-section, flux density}.
}

%------------------------------------------------------------------------------------------------------------------------------------
%------------------------------------------------------------------------------------------------------------------------------------
%------------------------------------------------------------------------------------------------------------------------------------

\section{Introduction}

Neutron moderation theory is an important issue in nuclear reactor 
physics~\cite{Weinberg1958, Akhiezer2002, Galanin1960, Feinberg1978, BartolomeyBat1989, Shirokov1998, Stacey2001, Vladimirov1986, Oka2014}. 
Conventional nuclear reactor physics theory of neutron moderation does
not consider temperature of the moderating medium, because it is based on
neutron scattering law not considering medium nuclei thermal motion. Within
this theory the neutron moderation spectrum is described by an analytic
expression known as Fermi moderation spectrum~\cite{Weinberg1958, Akhiezer2002, Galanin1960, Feinberg1978,
BartolomeyBat1989, Shirokov1998, Stacey2001, Vladimirov1986, Oka2014}. Fermi moderation spectrum also does not depend
on moderating medium temperature and therefore it cannot describe
experimental neutrons moderation spectrum correctly in low-energy range, e.g.
neutron spectra of thermal nuclear power reactors. Due to the fact that
experimental neutron spectra for thermal nuclear reactors are similar to
Maxwell distribution~\cite{Weinberg1958, Galanin1960, Stacey2001, Gurevich1965, Vlasov1971}, an assumption has
been made that neutron moderation spectrum can be described by this
distribution in low-energy range. This assumption enabled development of widely
used contemporary semi-empiric calculation algorithm for neutron moderation
spectra calculation. According to this algorithm the neutron moderation
spectrum is a combination of Fermi moderation spectrum at high energies and
the Maxwell's distribution at low energies. Hence it contains a formula to
determine neutron gas temperature based on moderating medium temperature which
determines low-energy spectrum range~\cite{Weinberg1958, Akhiezer2002, Galanin1960, Feinberg1978, BartolomeyBat1989, Shirokov1998,
Stacey2001, Vladimirov1986, Oka2014}. And the formula itself was obtained in the past by numerical
approximation of experimental spectra of several different thermal nuclear
reactors available at that moment~\cite{Weinberg1958}, and it is still widely used
in nuclear reactor physics (see e.g.~\cite{BartolomeyBat1989, Shirokov1998, Stacey2001, Vladimirov1986, Oka2014, RusovEnergies2011,
RusovVANT2011, RusovWJNST2013, RusovJMP2013, RusovKhariton2010}). Therefore in the strict sense there is no neutron
moderation theory available, and creating one is an essential task.

Let us also note, that the neutron moderation spectra in different
media can be modeled by Monte-Carlo approach. This method is the basis for
neutron moderation spectra calculation implemented in different reactor
computer algorithms. This includes software with the corresponding
interfaces designed to model the neutron transport, photons, etc in different
media by Monte-Carlo method, like MCNP4, GEANT4~\mbox{\cite{MCNP2000, GEANT42003}} and others.
They enable determination of neutron moderation spectrum by numerical
solution of kinematics problem of neutron scattering scattering, taking
into account a possible neutron absorption by moderating medium. This
computation is repeated multiple times for random neutron and nucleus initial
velocities to obtain resulting neutron moderation spectrum by averaging of the
accumulated results across all samples. But this is rather a modeling approach
than a theory.

Lack of neutron moderation theory first of all hinders investigation of emergency modes of nuclear reactors and development of new generation of reactors physics, e.g. soliton fission reactors~\cite{RusovSTNI2015, RusovPNE2015}, impulse reactors, neutron breeders (boosters), subcritical assemblies~\cite{Kolesov2006, Lukin2006, Arapov2010} and natural-occurring nuclear reactors, e.g. georeactor~\cite{RusovJGR2007}.

This paper proposes a further development of our recently published
work~\cite{Rusov2017} which provided a base theory for neutron moderation 
taking into account the temperature of the fissile medium. The base theory
model of neutron moderation does not fully consider inelastic scattering
reactions and therefore it is applicable to moderating media consisting of
nuclei with negligibly small inelastic scattering cross-section. The current
paper develops a neutron moderation theory for neutrons emitted by an isotropic
neutron source considering inelastic scattering. Account for inelastic
scattering reactions is especially important for moderating media consisting of
heavy nuclei. According to~\cite{Vlasov1971}, inelastic neutron scattering by
heavy nuclei manifests itself at neutron energy higher than several hundreds
keV, in contrast to the light nuclei, where inelastic scattering manifests
itself at energy higher than few MeV.

Our main interest is the physics of fissile neutron-multiplying media
(reactor fuel), and therefore the neutron moderation theory was developed first
of all to be applicable to such media. The assumption of neutron source
isotropy corresponds to a common way of introducing the neutron
source in fissile materials.

Any fission medium (nuclear fuel during the reactor operation, but in general
case any medium wherein a chain reaction takes place) is thermodynamically
unstable due to fission processes followed by release of high amounts of
energy and emission of neutrons and other particles. Moreover, the change in
nuclide composition, thermal conductivity and radiation defects dynamics may
sometimes lead to the medium deformation or even disruption. This way, the
fission medium of the reactor where fission processes take place is an open
physical system in non-equilibrium thermodynamical state. Such system may be
described within the framework of non-linear, non-equilibrium
thermodynamics of open physical systems. The non-equilibrium stationary states,
satisfying the Prigogine criterion of minimum entropy production, may exist in
such systems~\cite{Prigogine1968, Bakhareva1976, Kvasnikov2003}. From non-linear non-stationary
thermodynamics it is known that the emergence and the type of such
stationary mode depend not only on system internal parameters (internal
entropy) but also on boundary conditions (entropy flow at boundaries). For
example, implementation of a stationary mode in a non-equilibrium system
(so called non-equilibrium stationary state) demands constancy of boundary
conditions, e.g.~\cite{Bakhareva1976}.

In this paper the following simplifications are used for building the model of
neutron moderation process in a fissile medium. Two thermodynamic sub-systems
are distinguished: moderating neutrons sub-system and moderating medium nuclei
sub-system. These are the open physical systems interacting with one
another. According to the above argumentation, in reality both systems are
in non-equilibrium state. However, in our model we use a simplification that
the moderating medium nuclei sub-system is in state close to equilibrium due to
its inertness with respect to perturbations. Therefore by neglecting the
influence of neutrons on the nuclei sub-system, we assume that the latter is in
thermodynamic equilibrium. This simplification enables us to introduce the
moderating medium temperature into the model and to describe the nuclei kinetic
energy by Maxwell's distribution. We consider the moderating neutrons
sub-system in non-equilibrium state, so we do not introduce the temperature for
it in our model.

Let us emphasize that, as noted above, in traditional approach to deriving
neutron moderation spectrum a neutron gas sub-system temperature is introduced.
This implies a simplified assumption that neutrons sub-system is also in
thermodynamic equilibrium state which we avoided. Moreover, as noted above,
the formula for the neutron gas temperature is connected with the moderating medium
temperature through an empirical relation.

Let us note that the anisotropy of elastic and inelastic scattering
reactions was not considered, and this task is postponed for further theory
development together with the problem of accounting for moderating medium nuclei
interaction.

Let us also note that the previous history of this topic is given in~\mbox{\cite{Rusov2017}} in detail.

%------------------------------------------------------------------------------------------------------------------------------------
%------------------------------------------------------------------------------------------------------------------------------------
%------------------------------------------------------------------------------------------------------------------------------------

\section{Kinematics of inelastic neutron scattering at moderating medium nucleus}
\label{sect2}

If the neutron inelastic scattering is relevant to the medium (e.g. for
heavy nuclei such as uranium-238), the kinematics presented in our
paper~\cite{Rusov2017} should be somewhat refined.

According to~\cite{BartolomeyBat1989, Shirokov1998} neutron inelastic scattering reaction at a nucleus progresses through an excited compound nuclei state, i.e. at the first phase of the reaction the neutron moderating medium nucleus captures a neutron:
\begin{equation}
(A,Z)+ {}^1_0n \rightarrow (A+1,Z)^*
,
\label{eq1}
\end{equation}
where $(A,Z)$ is neutron moderating medium nucleus with mass number $A$ and charge $Z$; ${}^1_0n$ -- neutron (0 -- zero charge, 1 -- neutron mass number); $(A+1,Z)^*$ -- a compound nucleus with mass number $A+1$ and charge $Z$; ${}^*$ -- denotes nucleus excited state. During the next phase of the reaction the compound nucleus decomposes producing a nucleus of the neutron moderating medium in excited state and emitting a neutron. The nucleus is de-excited by emitting a $\gamma$-quantum:
\begin{equation}
(A+1,Z)^* \rightarrow (A,Z)^*+ {}^1_0n \rightarrow (A,Z)+ {}^1_0n +\gamma
.
\label{eq2}
\end{equation}

Here we should note that the compound nucleus decomposition is a two-particle
nuclear reaction, so the solution of the corresponding kinematics problem
is unambiguous (see e.g.~\mbox{\cite{Sitenko1993})}.

\begin{figure}
\includegraphics[width=3in]{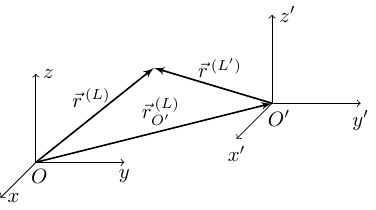}
\centering
\caption{$L$ and $L'$ laboratory coordinate systems.}
\label{fig1}
\end{figure}

Let us consider neutron inelastic scattering at a moderating medium nucleus.
Neutron moderating medium (following~\cite{Rusov2017}) is described within the
framework of gas model. I.e. it is assumed that the medium nuclei do not interact with one another, but possess kinetic energy due to thermal motion. Following~\cite{Rusov2017} Let us introduce two laboratory coordinates systems (fig.~\ref{fig1}):
\begin{itemize}
\item a stationary laboratory coordinate system, the $L$-system
\item a non-stationary laboratory coordinate system or $L'$-system that moves
relative to $L$-system with a constant velocity equal to the velocity of
thermal motion of the moderating medium nucleus at which the neutron is
scattered.
\end{itemize}

Let us note that here we consider a special case when the
orientation of $L$ and $L'$ coordinate systems is the same, and
the radius vector of $L'$-system origin in $L$-system coincides with the
moderator nucleus radius vector in $L$-system. So the moderator nucleus
initially rests in $L'$-system.

Here we use the same notation as in~\cite{Rusov2017}:
\\
$m_1 = m_{neutron}$ -- neutron mass;
\\
$m_2 = m_{nucleus}$ -- nucleus mass;
\\
$\vec{r}_1^{\,(L)}$ -- neutron radius vector in $L$-system;
\\
$\vec{r}_2^{\,(L)}$ -- nucleus radius vector in $L$-system;
\\
$\vec{r}_C^{\,(L)}$ -- inertia center radius vector in $L$-system;
\\
$\vec{r}_1^{\,(L')}$ -- neutron radius vector in $L'$-system;
\\
$\vec{r}_2^{\,(L')}$ -- nucleus radius vector in $L'$-system;
\\
$\vec{r}_C^{\,(L')}$ -- radius vector of inertia center in $L'$-system
\\
$\vec{V}_{10}^{\,(L)}$ -- neutron velocity before interaction with the nucleus in $L$-system;
\\
$\vec{V}_1^{\,(L)}$ -- neutron velocity after interaction with the nucleus in $L$-system;
\\
$\vec{V}_{20}^{\,(L)}$ -- nucleus velocity before interaction with the neutron in $L$-system;
\\
$\vec{V}_2^{\,(L)}$ -- nucleus velocity after interaction with the neutron in $L$-system;
\\
$\vec{V}_{10}^{\,(L')}$ -- neutron velocity before interaction with the nucleus in $L'$-system;
\\
$\vec{V}_1^{\,(L')}$ -- neutron velocity after interaction with the nucleus in $L'$-system;
\\
$\vec{V}_{20}^{\,(L')}$ -- nucleus velocity before interaction with the neutron in $L'$-system;
\\
$\vec{V}_2^{\,(L')}$ -- nucleus velocity after interaction with the neutron in $L'$-system;
\\
$\vec{V}_C^{\,(L')}$ -- inertia center velocity in $L'$-system.

The radius vectors of a point in $L$ and $L'$ systems are connected as
follows:
\begin{equation}
\vec{r}^{\,(L)}= \vec{r}_{O'}^{\,(L)} + \vec{r}^{\,(L')}
,
\label{eq3}
\end{equation}
where $\vec{r}_{O'}^{\,(L)}$ is the radius vector of $L'$-system
origin in $L$-system (see fig.~\ref{fig1}).

Therefore, according to the problem definition:
\begin{equation}
\vec{V}_{10}^{\,(L)} = \frac{\mathrm{d}\vec{r}_1^{\,(L)}}{\mathrm{d}t} \neq 0
~~~~~
\text{and}
~~~~~
\vec{V}_{20}^{\,(L)} = \frac{\mathrm{d}\vec{r}_2^{\,(L)}}{\mathrm{d}t} \neq 0
.
\label{eq4}
\end{equation}

In $L'$-system before the impact the nucleus rests, i.e.:
\begin{equation}
\vec{V}_{20}^{\,(L')} = 0
.
\label{eq5}
\end{equation}

The relation between coordinates of $m_1$ and $m_2$ in $L$ and
$L'$-systems is given by (\ref{eq3}). And the relation between velocities
(inertial systems velocity addition law, based on Galilean relativity principle):
\begin{equation}
\vec{V}_1^{\,(L')} = \vec{V}_1^{\,(L)} - \vec{V}_{20}^{\,(L)}
~~~~~
\text{and}
~~~~~
\vec{V}_2^{\,(L')} = \vec{V}_2^{\,(L)} - \vec{V}_{20}^{\,(L)}
.
\label{eq6}
\end{equation}

Due to the fact that the moderating medium nucleus rests in $L'$-system,
the solution of kinematic problem of inelastic scattering of the neutron at the
nucleus in $L'$-system may be expressed similar to the solution of
two-particle nuclear reaction kinematics with non-zero reaction thermal effect,
given in e.g.~\cite{Sitenko1993}. Indeed, according to the inelastic neutron
scattering reaction schema given by (\ref{eq1}) and (\ref{eq2}), the thermal
effect of such reaction equals to emitted $\gamma$-quantum energy,
further denoted by $E_\gamma$. Next, let us consider our problem in analogy
with~\cite{Sitenko1993}.

It is convenient to solve a two-particle interaction problem in a
center-of-mass frame, i.e. in $C$-system.

For inelastic neutron scattering reaction in $C$-system the total isolated
system momentum conservation law is valid. However, the total kinetic energy
conservation law in this isolated system does not work.

From momentum conservation law for two impacting particles in $C$-system we obtain:
\begin{equation}
\vec{P}_{10}^{\,(C)}+ \vec{P}_{20}^{\,(C)} = \vec{P}_1^{\,(C)} + \vec{P}_2^{\,(C)} = 0
,
\label{eq7}
\end{equation}
where:
\\
$\vec{P}_{10}^{\,(C)}$ -- neutron momentum before interaction with the
nucleus in $C$-system;
\\
$\vec{P}_1^{\,(C)}$ -- neutron momentum after interaction with the nucleus
in $C$-system;
\\
$\vec{P}_{20}^{\,(C)}$ -- nucleus momentum before interaction with the
neutron in $C$-system;
\\
$\vec{P}_2^{\,(C)}$ -- nucleus momentum after interaction with the
neutron in $C$-system;

As follows from relations (\ref{eq7}):
\begin{equation}
m_1 \cdot \vec{V}_{10}^{\,(C)} = - m_2 \cdot \vec{V}_{20}^{\,(C)}
~~~~~
\text{and}
~~~~~
m_1 \cdot \vec{V}_1^{\,(C)} = - m_2 \cdot \vec{V}_2^{\,(C)}
.
\label{eq8}
\end{equation}

Velocity modulus can be obtained from (\ref{eq8}):
\begin{equation}
\left\lvert \vec{V}_{10}^{\,(C)} \right\rvert = {V_{10}}^{\,(C)} = \frac{m_2}{m_1} \left\lvert \vec{V}_{20}^{\,(C)} \right\rvert = \frac{m_2}{m_1} {V_{20}}^{\,(C)}
\label{eq9a}
\end{equation}
and
\begin{equation}
\left\lvert \vec{V}_1^{\,(C)} \right\rvert = {V_1}^{\,(C)} = \frac{m_2}{m_1} \left\lvert \vec{V}_2^{\,(C)} \right\rvert = \frac{m_2}{m_1} {V_2}^{\,(C)}
.
\label{eq9b}
\end{equation}

By introducing neutron $A_{neutron}=1$ and nucleus $A_{nucleus}=A$ mass numbers and assuming $m_1={m_{neutron}} \approx A_{neutron} = 1$ and $m_2 = m_{nucleus} \approx A_{nucleus} = A$ we obtain the following expressions for (\ref{eq8}) and (\ref{eq9a})-(\ref{eq9b}):
\begin{equation}
\vec{V}_{10}^{\,(C)} = - A \cdot \vec{V}_{20}^{\,(C)}
~~~~~
\text{and}
~~~~~
\vec{V}_1^{\,(C)} = - A \cdot \vec{V}_2^{\,(C)}
,
\label{eq10}
\end{equation}
and
\begin{equation}
{V_{10}}^{\,(C)} = + A \cdot V_{20}^{\,(C)}
~~~~~
\text{and}
~~~~~
{V_1}^{\,(C)} = + A \cdot V_2^{\,(C)}
.
\label{eq11}
\end{equation}

The moderating medium nucleus and neutron inertia center coordinates may be expressed in the following way:
\begin{equation}
\vec{r}_C^{\,(L')} = (1 \cdot \vec{r}_1^{\,(L')} + A \cdot \vec{r}_2^{\,(L')} ) \cdot \frac{1}{A+1}
.
\label{eq12}
\end{equation}

Considering the fact, that in $L'$-system the initial nucleus velocity prior to impact is $\vec{V}_{20}^{\,(L')} = 0$, the inertia center velocity for an isolated system consisting of two particles (i.e. the neutron and the nucleus) in $L'$-system is the following:
\begin{equation}
\vec{V}_C^{\,(L')} = \frac{1}{A+1} \cdot \vec{V}_{10}^{\,(L')}
.
\label{eq13}
\end{equation}

Due to total momentum conservation law the inertia center velocity in $L'$-system doesn't change after the impact. Therefore we omit corresponding indexes denoting values prior and after the impact for the inertia center velocity.

Due to the fact that inertia center coordinate system ($C$-system) moves relative to laboratory system with velocity of inertia center in $L'$-system, the neutron velocity prior to impact in $C$-system is:
\begin{equation}
\vec{V}_{10}^{\,(C)} = \vec{V}_{10}^{\,(L')} - \vec{V}_C^{\,(L')}
.
\label{eq14}
\end{equation}

Substituting it into (\ref{eq13}) we obtain:
\begin{equation}
\vec{V}_{10}^{\,(C)} = \vec{V}_{10}^{\,(L')}-\frac{1}{A+1} \cdot \vec{V}_{10}^{\,(L')} = \frac {A}{A+1} \cdot \vec{V}_{10}^{\,(L')}
.
\label{eq15}
\end{equation}

Using (\ref{eq8}) and considering (\ref{eq13}), we find the nucleus velocity in $C$-system prior to impact:
\begin{equation}
\vec{V}_{20}^{\,(C)} = - \frac {1}{A+1} \cdot \vec{V}_{10}^{\,(L')}
.
\label{eq16}
\end{equation}

The total kinetic energy of the system (two-particle system consisting of $(A,Z)^*$ and ${}^1_0n$) after the reaction $T^{(C)}$  equals to:
\begin{equation}
T^{(C)} = T_0^{(C)} + E_\gamma
,
\label{eq19}
\end{equation}
where $T^{(C)}_0$ -- total kinetic energy of the system before the reaction.

Using (\ref{eq15}) and (\ref{eq16}), we can express the total kinetic energy of the system prior to reaction:
\begin{equation}
T_0^{(C)} =
\frac{\left\lvert \vec{P}_{10}^{\,(C)} \right\rvert ^2 } {2 \mu} =
\frac{\mu \left( V_{10}^{(L')} \right) ^2 } {2}
,
\label{eq20}
\end{equation}
where $\mu=A/(A+1)$ -- reduced mass of the particles prior to reaction.

Then kinetic energy of the system after the reaction according to (\ref{eq1}) and (\ref{eq2}) equals:
\begin{equation}
T^{(C)} =
\frac{\left\lvert \vec{P}_1^{\,(C)} \right\rvert ^2 } {2 \mu} =
\frac{\left\lvert \vec{P}_{10}^{\,(C)} \right\rvert ^2 } {2 \mu} - E_\gamma  =
\frac{\mu \left( V_{10}^{(L')} \right) ^2 } {2} - E_\gamma
.
\label{eq21}
\end{equation}

The neutron momentum modulus after the reaction can be expressed based on (\ref{eq21}):
\begin{align}
\left\lvert \vec{P}_1^{\,(C)} \right\rvert & =
\sqrt{2 \mu' T^{\,(C)}} =
\sqrt{2 \mu' \left(T_0^{\,(C)} - E_\gamma\right)} = \nonumber \\
& = \sqrt{2 \mu' \left[ \frac {\mu \left( V_{10}^{\,(L')} \right)^2 } {2} - E_\gamma \right]}
,
\label{eq22}
\end{align}
where $\mu' = (A+E_\gamma / c^2) / (A+1)$ -- reduced mass of the particles after the reaction, where $c$ is the speed of light in vacuum.

Denoting a unit vector along the neutron velocity in $C$-system after the reaction as $\vec{e}_1$, according to (\ref{eq22}) we obtain:
\begin{equation}
\vec{V}_1^{\,(C)} =
\sqrt{2 \mu' \left[ \frac {\mu \left( V_{10}^{\,(L')} \right)^2 } {2} - E_\gamma \right]} \cdot \vec{e}_1
.
\label{eq23}
\end{equation}

Let us note that the angular distribution of inelastic neutron scattering
may be almost always considered spherically-symmetric~\cite{BartolomeyBat1989, Stacey2001} just as
for the elastic scattering.

The neutron velocity vector after the reaction in $L'$-system may be obtained by adding neutron velocity vector after impact in $C$-system (\ref{eq23}) and velocity vector in $C$-system (\ref{eq13}):
\begin{equation}
\vec{V}_1^{\,(L')} = \vec{V}_1^{\,(C)}+\frac{1}{A+1} \vec{V}_{10}^{\,(L')}
.
\label{eq24}
\end{equation}

\begin{figure}
\includegraphics[width=2.5in]{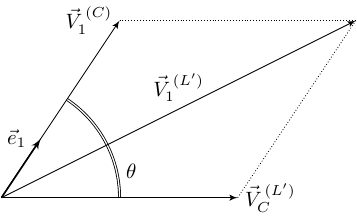}
\centering
\caption{Velocity parallelogram in $L'$-system after the impact.}
\label{fig2}
\end{figure}

Based on velocity parallelogram shown in fig.~\ref{fig2}, square of neutron velocity modulus in $L'$-system after the impact is:
\begin{align}
\left( \vec{V}_1^{\,(L')} \right)^2 & =
\left( \vec{V}_1^{\,(C)} \right)^2 +
\left( V_{10}^{\,(L')}\cdot \frac{1}{A+1} \right)^2 + \nonumber \\
& + \frac { 2 \cdot \left( V_1^{\,(C)} \right) \cdot \left( V_{10}^{\,(L')} \right) } {A+1} \cdot
\cos \theta
,
\label{eq25}
\end{align}
where $\theta$ is neutron exit angle in $C$-system (fig.~\ref{fig2}).

The energy of the emitted $\gamma$-quantum, $E_\gamma$, is confined within
the range:
\begin{equation}
0 \leq E_\gamma \leq \frac{A}{A+1} \cdot \frac{ \left( \vec{V}_{10}^{\,(L')} \right)^2 } {2} < E_f
,
\label{eq26}
\end{equation}
where $E_f$ -- is the margin neutron energy, that causes fission of the compound nucleus $(A+1,Z)^*$.

As follows from neutron and nucleus (at rest in $L'$-system) isolated system total energy conservation law, the compound nucleus excitation energy equals to (see e.g.~\cite{BartolomeyBat1989, Shirokov1998}):
\begin{equation}
E^* = \varepsilon_n + \frac{A}{A+1}T_n^{(L')}
,
\label{eq27}
\end{equation}
where $\varepsilon_n$ -- neutron bond energy in the compound nucleus, $T_n^{(L')}$ -- neutron kinetic energy prior to scattering in $L'$-system.

During decomposition of the compound nucleus into a moderating medium nucleus in a non-excited state, neutron and $\gamma$-quantum (see reaction schema (\ref{eq2})) a fraction of the excited nucleus energy $\varepsilon_n$ denotes the work against strong interaction forces during neutron emission, and the remaining excitation energy fraction equals to sum of system kinetic energy (nucleus and neutron) in $C$-system after scattering and $\gamma$-quantum energy:
\begin{equation}
\frac {A}{A+1} T_n^{(L')}= \frac{\left( \vec{P}_1^{\,(C)} \right)^2} {2{\mu'}} + E_\gamma
.
\label{eq28}
\end{equation}

A stochastic multitude of reactions of neutrons inelastic scattering at moderating medium nuclei contribute to total neutron moderation spectrum, herewith the $\gamma$-quantum kinetic energy (and also square of the neutron momentum after scattering in $C$-system univocally related to it through (\ref{eq28})) varies randomly within the energy range determined by (\ref{eq26}). There are no reasons to suggest that some value of $\gamma$-quantum kinetic energy is randomly preferred, i.e. that a $gamma$-quantum with some specific energy is somehow emitted more frequently than others. Therefore it may be assumed that a random $\gamma$-quants energy value within the inelastic scattering of a multitude of neutrons at nuclei of the moderating medium will be given by equally probable distribution and the probability density is determined by the following expression:
\begin{equation}
\rho (E_\gamma) = \rho \left( \vec{P}_1^{\,(C)} \right) = \frac {1} { \left( \dfrac{A}{A+1} \right) \dfrac {\left( V_{10}^{\,(L')} \right)^2}{2}}
.
\label{eq29}
\end{equation}

The average value of square of neutron momentum and average value of square of neutron velocity after scattering in $C$-system (which are equal because the neutron has $A_{neutron}=1$) in case of equiprobable random distribution can be easily expressed as a sum of its maximal (at $E_\gamma = 0$) and minimal (at $E_\gamma = \frac{A}{A+1} T_n^{(L')}$) values divided by 2:

\begin{align}
& \left( \overline{V}_1^{\,(C)} \right)^2 =
\frac {1}{2} \left( 2 \left(\frac{A}{A+1}\right)^2 \frac{\left(\vec{V}_{10}^{\,(L')}\right)^2}{2} +0  \right) =
\nonumber \\ &=
\frac{1}{2} \left(\frac{A}{A+1}\right)^2 \left(\vec{V}_{10}^{\,(L')}\right)^2 =
\frac{A^2 B^2}{(A+1)^2}{\left(\vec{V}_{10}^{\,(L')}\right)^2} ,
\label{eq30}
\end{align}

\begin{equation}
B^2 = \frac {1}{2} .
\label{eq31}
\end{equation}

Knowing average value of neutron velocity modulus after scattering in $C$-system, we can analogous to expression (\ref{eq23}) multiply it by unit vector of neutron exit direction $\vec{e}_1$ in $C$-system and obtain an expression for average neutron velocity vector after scattering:
\begin{equation}
\vec{\overline{V}}_1^{\,(C)} =
\frac{AB}{A+1} V_{10}^{\,(L')} \cdot \vec{e}_1
.
\label{eq32}
\end{equation}

Therefore, the average value of square of neutron velocity after scattering in $L'$-system can be obtained analogous to (\ref{eq32}), i.e.:
\begin{align}
\left( \vec{\overline{V}}_1^{\,(L')} \right)^2 & =
\left( \vec{\overline{V}}_1^{\,(C)} \right)^2 +
\left( \vec{V}_{10}^{\,(L')} \cdot \frac{1}{A+1} \right)^2 + \nonumber \\
& + \frac {2 \cdot \left( \overline{V}_1^{\,(C)} \right) \cdot \left( V_{10}^{\,(L')} \right)}
{A+1} \cdot \cos \theta
.
\label{eq33}
\end{align}

Substituting $\overline{V}_1^{\,(C)}$ from (\ref{eq32}) into (\ref{eq33}) we obtain the following expression:
\begin{equation}
\left(\overline{V}_1^{\,(L')} \right)^2 =
\frac {\left(V_{10}^{\,(L')} \right)^2 \cdot
\left( A^2 B^2 + 2AB\cos \theta +1 \right)^2} {(A+1)^2}
.
\label{eq34}
\end{equation}

The ratio of square of average neutron velocity after inelastic scattering $\left(\overline{V}_1^{\,(L')} \right)^2$ to square of neutron velocity prior to scattering $\left(V_{10}^{\,(L')} \right)^2$ in $L'$-system can be derived from (\ref{eq34}) considering inelastic scattering of multitude of neutrons that have equal velocity module prior to scattering. This ratio is also equal to ratio of average neutron kinetic energy after scattering $\overline{E}_2^{\,(L')}$ to neutron kinetic energy prior to scattering $E_1^{\,(L')}$:
   
\begin{equation}
\frac { \left(\overline{V}_1^{\,(L')} \right)^2 } {\left(V_{10}^{\,(L')} \right)^2 } =
\frac {\overline{E}_2^{\,(L')}} {E_1^{\,(L')}} =
\frac { A^2B^2 + 2AB\cos\theta +1 }{(A+1)^2}
.
\label{eq35}
\end{equation}

By introducing inelastic scattering coefficients in the following way:
\begin{equation}
\widetilde{\alpha}_1^2 = \left( \frac {AB-1}{AB+1} \right)^2
~~~~~
\text{and}
~~~~~
\widetilde{\alpha}_2^2 = \left( \frac {AB+1}{A+1} \right)^2
,
\label{eq36}
\end{equation}
expression (\ref{eq35}) transforms into:
\begin{equation}
\frac {\overline{E}_2^{\,(L')}} {E_1^{\,(L')}} =
\frac{\widetilde{\alpha}_2^2}{2} \left[ (1+\widetilde{\alpha}_1) + (1-\widetilde{\alpha}_1)\cos\theta \right]
.
\label{eq37}
\end{equation}

Now, expressing the vectors of average neutron velocity after scattering in $C$-system through analogous vector of average neutron velocity after scattering in $L$-system, we can rewrite (\ref{eq37}) in the following form:
\begin{align}
& \frac { \left( \vec{\overline{V}}_1^{\,(L)} - \vec{V}_{20}^{\,(L)} \right)^2 }
{ \left( \vec{V}_{10}^{\,(L)} - \vec{V}_{20}^{\,(L)} \right)^2 }  =
\frac {\overline{E}_2^{\,(L')}} {E_1^{\,(L')}} = \nonumber \\
& =\frac{\widetilde{\alpha}_2^2}{2} \left[ (1+\widetilde{\alpha}_1) + (1-\widetilde{\alpha}_1)\cos\theta \right]
. 
\label{eq38}
\end{align}

From expression (\ref{eq38}) for inelastic scattering we obtain an expression analogous to elastic scattering expression obtained in~\cite{Rusov2017}:
\begin{align}
& \frac{\left( \vec{\overline{V}}_1^{\,(L)} \right)^2}
     {\left( \vec{V}_{10}^{\,(L)} \right)^2} =
\frac{ \overline{E}_1^{\,(L)} } { E_{10}^{\,(L)} } = 
\frac{\widetilde{\alpha}_2^2}{2}
\left[ (1+\widetilde{\alpha}_1)+(1-\widetilde{\alpha}_1)\cos\theta \right] \times \nonumber \\
& \times \left[
1-2 \frac{V_{10}^{\,(L)} V_{20}^{\,(L)} \cos\beta} 
{\left( V_{10}^{\,(L)} \right)^2 }
+ \frac{ \left(V_{20}^{\,(L)}\right)^2 }{ \left(V_{10}^{\,(L)}\right)^2 }
\right]+ \nonumber \\
&+
2 \frac{ \left( \vec{\overline{V}}_1^{\,(L)} , \vec{V}_{20}^{\,(L)} \right) }
{\left(V_{10}^{\,(L)}\right)^2 } -
\frac{\left(V_{20}^{\,(L)}\right)^2 }
{\left(V_{10}^{\,(L)}\right)^2 } = \nonumber \\
&=
\frac{\widetilde{\alpha}_2^2}{2} \left[ (1+\widetilde{\alpha}_1)+(1-\widetilde{\alpha}_1)\cos\theta \right] - \nonumber \\
& - \frac{\widetilde{\alpha}_2^2}{2} \left[ (1+\widetilde{\alpha}_1)+(1-\widetilde{\alpha}_1)\cos\theta \right]
\frac {V_{10}^{\,(L)}V_{20}^{\,(L)}\cos\beta}
{\left(V_{10}^{\,(L)}\right)^2 } + \nonumber \\
&+
2 \frac {\overline{V}_1^{\,(L)}V_{20}^{\,(L)}\cos\gamma}
{\left(V_{10}^{\,(L)}\right)^2 } - \nonumber \\
&-
\left\{
1 -
\frac{\widetilde{\alpha}_2^2}{2} \left[ (1+\widetilde{\alpha}_1)+(1-\widetilde{\alpha}_1)\cos\theta \right]
\right\}
\frac {1 \cdot E_{20}^{\,(L)}} {A \cdot E_{10}^{\,(L)}}
,
\label{eq39}
\end{align}

\noindent
where $\overline{E}_1^{\,(L)}=\frac{1}{2}\left( \vec{\overline{V}}_1^{\,(L)} \right)^2$; $\cos \beta$ -- cosine of the angle between $\vec{V}_{10}^{\,(L)}$ and $\vec{V}_{20}^{\,(L)}$, expressed through scalar product $\left( \vec{V}_{10}^{\,(L)} , \vec{V}_{20}^{\,(L)} \right)$ in the following way:
\begin{equation}
\cos \beta =
\frac {\left( \vec{V}_{10}^{\,(L)} , \vec{V}_{20}^{\,(L)} \right)}
{\left\lvert \vec{V}_{10}^{\,(L)} \right\rvert \left\lvert \vec{V}_{20}^{\,(L)} \right\rvert} =
\frac {\left( \vec{V}_{10}^{\,(L)} , \vec{V}_{20}^{\,(L)} \right)}
{V_{10}^{\,(L)} V_{20}^{\,(L)}}
,
\label{eq40}
\end{equation}
$\cos \gamma$ -- cosine of the angle between $\vec{\overline{V}}_1^{\,(L)}$ and $\vec{V}_{20}^{\,(L)}$:
\begin{equation}
\cos \gamma =
\frac {\left( \vec{\overline{V}}_1^{\,(L)} , \vec{V}_{20}^{\,(L)} \right)}
{\left\lvert \vec{\overline{V}}_1^{\,(L)} \right\rvert \left\lvert \vec{V}_{20}^{\,(L)} \right\rvert}
=
\frac {\left( \vec{\overline{V}}_1^{\,(L)} , \vec{V}_{20}^{\,(L)} \right)}
{\overline{V}_{1}^{\,(L)} V_{20}^{\,(L)}}
.
\label{eq41}
\end{equation}

As is be shown below, for our purposes it is sufficient to limit oneself to
expression (\ref{eq39}) without performing the rest of the transformations of
the Eq.~(\ref{eq39})-(\ref{eq41}) and obtaining the exact solution of the
considered kinematic problem. It would only yield a set of lengthy expressions,
since the intermediate solution of the problem (\ref{eq39}) includes the angles
between the neutron and nucleus velocity vectors (\ref{eq40}) and (\ref{eq41}).
These cosines would also require the adaptation to the $L$-system. Therefore
reducing the solution to a single analytic expression is pointless within
the present paper.

%------------------------------------------------------------------------------------------------------------------------------------
%------------------------------------------------------------------------------------------------------------------------------------
%------------------------------------------------------------------------------------------------------------------------------------

\section{Neutron inelastic scattering law considering moderator nuclei thermal motion}
According to the kinematics of the neutron inelastic scattering on a moderator
nucleus given in Section~\mbox{\ref{sect2}} by (\mbox{\ref{eq39}}), the probability density
function for scattering to a certain energy interval is determined by several
independent random variables ($\theta$, $\beta$, $\gamma$, $E_N$). Thus:

\begin{equation}
P \left(\theta, \beta, \gamma, E_N^{\,(L)} \right) =
P_{\theta}(\theta) \cdot
P_{\beta}(\beta) \cdot
P_{\gamma}(\gamma) \cdot
P_{E_N} \left( E_N^{\,(L)} \right) 
\label{eq42}
\end{equation}

As known in nuclear physics, as long as the compound nucleus decomposition  does not depend on history of its formation, the inelastic scattering of neutrons in inertia center coordinate system is spherically symmetric (isotropic). Therefore for  $P(\theta)\mathrm{d}\theta$ we obtain:
\begin{align}
P(\theta)\mathrm{d}\theta & =
\int\limits_0^{2\pi} {\left[ P(\theta,\phi) \right] \mathrm{d}\phi} =
\int\limits_0^{2\pi} {\frac{r \sin\theta \mathrm{d}\phi \cdot r \mathrm{d}\theta}{4\pi r^2}} = \nonumber \\
& = 
\frac{\sin{\theta} \mathrm{d}\theta}{4\pi} \int\limits_0^{2\pi}\mathrm{d}\phi =
\frac{1}{2}\sin\theta \mathrm{d}\theta
,
\label{eq43}
\end{align}
where $\phi$ denotes azimuth angle of spherical coordinates $r$, $\theta$, $\phi$, introduced in inertia center coordinates system.

Due to the fact that thermal motion of the moderating medium nuclei is chaotic and the neutron source is isotropic (neutron source emits a neutron group with given energy and isotropic spatial distribution of their velocity vectors directions), the distribution of velocity vectors directions in space for neutrons after inelastic scattering is given by equiprobable random distribution law along $\beta$ and $\gamma$ angles as a part of (\ref{eq39}). I.e. it is also spherically symmetric (isotropic). Therefore analogous to the previous case we obtain:
\begin{equation}
P(\beta) \mathrm{d}\beta = \frac{1}{2} \sin \beta \mathrm{d}\beta
,
\label{eq44}
\end{equation}
\begin{equation}
P(\gamma) \mathrm{d}\gamma = \frac{1}{2} \sin \gamma \mathrm{d}\gamma
.
\label{eq45}
\end{equation}

Let us average kinetic energy of the neutron after inelastic scattering at a nucleus (given by (\ref{eq39})) across spherically-symmetric distribution of thermal motion velocities of the moderating medium nuclei the and across isotropic neutron source (across isotropic spatial distribution of neutron velocity vectors with given energy, emitted by the neutron source). Sequentially we obtain the following expression:
\begin{align}
&\overline{\overline{E}}_1^{\,(L)} =
\int\limits_0^\pi {\int\limits_0^\pi {\overline{E}_1^{\,(L)}P(\beta)\mathrm{d}\beta P(\gamma) \mathrm{d} \gamma}} = \nonumber \\ 
&=
\overline{E}_{10}^{\,(L)}
\left\{
\frac{\widetilde{\alpha}_2^2}{2} \left[ (1+\widetilde{\alpha}_1)+(1-\widetilde{\alpha}_1)\cos\theta \right] - \right. \nonumber \\
&\left. -
\left[
1- \frac{\widetilde{\alpha}_2^2}{2}
\left[
(1+\widetilde{\alpha}_1)+(1-\widetilde{\alpha}_1)\cos\theta 
\right]
\frac{E_N^{\,(L)}}{A \cdot \overline{E}_{10}^{\,(L)}}
\right]
\right\}
.
\label{eq46}
\end{align}

Here $\overline{E}_{10}^{\,(L)}$ -- neutron energy for isotropic neutron source averaged across neutron momentum directions, which is equal to energy of neutrons emitted by the source $E_{10}^{(L)}$, i.e. $\overline{E}_{10}^{\,(L)} = E_{10}^{\,(L)}$. And $E_N^{\,(L)}$ is nucleus kinetic enerty, determined by Maxwell's distribution for neutron moderating medium nuclei~\cite{Levich1971}, which depends on temperature of the moderating medium as a parameter and has the following form:

\begin{equation}
P \left( E_N^{\,(L)} \right) \mathrm{d} E_N^{\,(L)} =
\frac {2}{\sqrt{\pi(kT)^3}} \exp{\left[-\frac { E_N^{\,(L)}} {kT}\right]}
\sqrt{ E_N^{\,(L)} } \mathrm{d} E_N^{\,(L)}
.
\label{eq47}
\end{equation}

Let us average the expression (\ref{eq46}) across Maxwell's distribution of the neutron moderating medium nuclei thermal motion (\ref{eq47}), considering $\overline{E}_{10}^{\,(L)} = E_{10}^{\,(L)}$ and using known expression~\cite{Levich1971}
\begin{equation}
\overline{E}_N^{\,(L)} = \int\limits_0^\infty E_N^{\,(L)} P\left( E_N^{\,(L)} \right) \mathrm{d}E_N^{\,(L)} = \frac{3}{2}kT
,
\end{equation}
we obtain the following:
\begin{align}
&\overline{\overline{\overline{E}}}_1^{\,(L)} = 
E_{10}^{\,(L)}
\left\{
\frac{\widetilde{\alpha}_2^2}{2} \left[ (1+\widetilde{\alpha}_1)+(1-\widetilde{\alpha}_1)\cos\theta \right] \right.
- \\
& -\left.
\left[
1-\frac{\widetilde{\alpha}_2^2}{2} \left[ (1+\widetilde{\alpha}_1)+(1-\widetilde{\alpha}_1)\cos\theta
\right]
\frac {\frac{3}{2}kT} {A \cdot E_{10}^{\,(L)}}
\right]
\right\}
. \nonumber
\label{eq48}
\end{align}

Since the functional relationship between 
$\overline{\overline{\overline{E}}}_1^{\,(L)}$ and $\theta$ is unique, as
follows from~(\ref{eq48}), the probability
$P \left(\overline{\overline{\overline{E}}}_1^{\,(L)}\right) \mathrm{d}E_1^{\,(L)}$
or the neutron with a
kinetic energy $E_{10}^{(L)}$ in the $L$-system before scattering to possess
the kinetic energy in the range from
$\overline{\overline{\overline{E}}}_1^{\,(L)}$ to
$\overline{\overline{\overline{E}}}_1^{\,(L)} + \mathrm{d}E_1^{\,(L)}$
after the scattering on the chaotically
moving moderator nuclei is determined by the $P(\theta) d\theta$
distribution~(\ref{eq43}).  Therefore we obtain the following relation (here we
omit the symbols of averaging and the laboratory coordinate system $L$ for
simplicity), i.e.

\begin{align}
& P(E_1) =
P(\theta) \left\lvert \frac{\mathrm{d}\theta}{\mathrm{d}E_1} \right\rvert = 
\frac{1}{2} \sin\theta \times \\
&\left\lvert
\frac{1}{
E_{10}^{\,(L)}
\left[
\dfrac{\widetilde{\alpha}_2^2}{2} (1-\widetilde{\alpha}_1) \sin\theta +
\dfrac{\widetilde{\alpha}_2^2}{2} (1-\widetilde{\alpha}_1) \sin\theta \dfrac {\frac{3}{2}kT} {A \cdot E_{10}^{\,(L)}}
\right]
}
\right\rvert
\nonumber  \\ 
&=
\frac{1} {
\left[ E_{10}^{\,(L)} + \frac{1}{A}\cdot\frac{3}{2}kT \right]
\widetilde{\alpha}_2^2(1-\widetilde{\alpha}_1)}
. \nonumber 
\label{eq49}
\end{align}

Therefore the obtained neutron inelastic scattering law considering thermal motion of moderating medium nuclei is the following:
\begin{equation}
\begin{cases}
\begin{split}
& P(E_1)\mathrm{d}E_1 =
\frac {\mathrm{d}E_1}
{\left[ E_{10}^{\,(L)} + \frac{1}{A}\cdot \frac{3}{2}kT \right]
\widetilde{\alpha}_2^2 (1-\widetilde{\alpha}_1)}
, \\
&~~~\text{in case}~~
{\scriptstyle
\widetilde{\alpha}_2^2\widetilde{\alpha}_1
\left( E_{10}^{\,(L)}+ \frac{1}{A}\cdot \frac{3}{2}kT \right)
\leq E_1 \leq
\left( E_{10}^{\,(L)}+ \frac{1}{A}\cdot \frac{3}{2}kT \right)
}
;
%\end{split}
\\
%\begin{split}
& P(E_1)   = 0
, \\
&~~~\text{in case}~~
{\scriptstyle
E_1 <
\widetilde{\alpha}_2^2\widetilde{\alpha}_1
\left( E_{10}^{\,(L)}+ \frac{1}{A}\cdot \frac{3}{2}kT \right)
%~\text{or}~
;~
E_1 >
\left( E_{10}^{\,(L)}+ \frac{1}{A}\cdot \frac{3}{2}kT \right)
}
.
\end{split}
\end{cases}
\label{eq50}
\end{equation}

In the conclusion of the section Let us stress that the law of inelastic scattering  (\ref{eq50}) is given for averaged neutron energy after scattering $E_1$. The neutron energy averaging is made across thermal (chaotic) motion of the neutron moderating medium nuclei and across isotropic neutron source. As follows from the scattering law (\ref{eq50}) all the neutrons emitted by the isotropic neutron source and having energy $E_{10}^{(L)}$ prior to scattering at the moderating medium nuclei, after inelastic scattering at moderating medium nuclei will have energy $E_1$ averaged across thermal (chaotic) motion of the moderating medium nuclei and across isotropic neutron source with probability given by (\ref{eq50}).

As will be shown below, the obtained scattering law in form (\ref{eq50})
enables obtaining expressions for neutron flux density and neutron energy
spectra in different moderators as a function of temperature.

%------------------------------------------------------------------------------------------------------------------------------------
%------------------------------------------------------------------------------------------------------------------------------------
%------------------------------------------------------------------------------------------------------------------------------------

\section{Neutron moderation in absorbing media containing several sorts of nuclei}

According to~\cite{Rusov2017}, an analytic solution for stationary balance
equation for moderated neutrons may be found as neutron flux density. In that
case the
% expression for the elastically scattered moderating neutron flux
%density is:
the elastically scattered neutron flux is given in Eq.~(\ref{eq51}),
where $Q(E)$ is the quantity of neutrons generated with energy $E$ per unit volume per unit time, $\Sigma_{el}$ -- total macroscopic cross section of elastic scattering of the moderating medium ($\Sigma_{el} = \sum\limits_i {\Sigma_{el}^i}$, where $\Sigma_{el}^i$ -- macroscopic cross section of the i-th nuclide in the moderator medium compound), $\Sigma_t = \Sigma_s + \Sigma_a$ -- total macroscopic cross section, $\Sigma_s$ -- total fission environment macroscopic cross section, $\Sigma_a$ -- total neutron absorption macroscopic cross section, $\left\lvert \overline{\xi}_{el} \right\rvert$ -- modulus of average-logarithmic energy decrement for elastic scattering $\xi_{el}$, which is determined in analogy to standard neutron moderation theory (see e.g.~\cite{Feinberg1978, BartolomeyBat1989, Shirokov1998, Rusov2017}), but through a new neutron elastic scattering law in~\cite{Rusov2017}. Let us recall that $\Sigma_s = \Sigma_{el}+\Sigma_{in}$, where $\Sigma_{el} = \Sigma_p + \Sigma_{rs}$ -- total macroscopic cross section of elastic scattering, $\Sigma_p$ -- total macroscopic cross section of potential scattering, $\Sigma_{rs}$ -- total macroscopic cross section of resonant scattering, $\Sigma_{in}$ -- inelastic scattering macroscopic cross section, see e.g.~\cite{Shirokov1998}.

\begin{equation}
\Phi_{el}(E) = 
\frac {\int\limits_E^\infty {\dfrac{\Sigma_{el}(E')}{\Sigma_t (E')} Q(E') 
\cdot 
\exp{ \left[
- \int\limits_{E}^{E'} {\frac {\Sigma_a (E'') \mathrm{d}E'' }
{\left\lvert \overline{\xi}_{el} \right\rvert \Sigma_t (E'')
\left[
E'' + \frac{1}{A}\cdot \frac{3}{2}kT
\right]}}
\right]}
\mathrm{d}E'}}
{\left[
E + \frac{1}{A}\cdot \frac{3}{2}kT \right]
\Sigma_t(E) \left\lvert \overline{\xi}_{el} \right\rvert}
+
\frac{ \dfrac{\Sigma_{el}(E)}{\Sigma_t (E)} Q(E) } {\Sigma_t (E)} ,
\label{eq51}
\end{equation}

The expression (\ref{eq51}) contains probability function of resonant neutron non-absorption~\cite{Feinberg1978, BartolomeyBat1989, Shirokov1998, RusovVANT2012, Rusov2017}, now containing moderating medium temperature:
\begin{equation}
\varphi(E) = \exp {
\left[
-
\int\limits_E^\infty {
\frac {\Sigma_a (E') \mathrm{d}E'}
{\left\lvert \overline{\xi}_{el} \right\rvert \Sigma_t (E')
\left[ E' + \frac{1}{A}\cdot \frac{3}{2}kT \right]}
}
\right]}
.
\label{eq52a}
\end{equation}

Given the inelastic scattering law (\ref{eq50}) and performing calculations analogous to the one made for elastic scattering in~\cite{Rusov2017}, we obtain an expression for inelastically scattered moderating neutrons flux density in the following form:

\begin{equation}
\Phi_{in}(E) = 
\frac {\int\limits_E^\infty {\dfrac{\Sigma_{in}(E')}{\Sigma_t (E')} Q(E') 
\cdot 
\exp{ \left[
- \int\limits_{E}^{E'} {\frac {\Sigma_a (E'') \mathrm{d}E'' }
{\left\lvert \overline{\xi}_{in} \right\rvert \Sigma_t (E'')
\left[
E'' + \frac{1}{A}\cdot \frac{3}{2}kT
\right]}}
\right]}
\mathrm{d}E'}}
{\left[
E + \frac{1}{A}\cdot \frac{3}{2}kT \right]
\Sigma_t(E) \left\lvert \overline{\xi}_{in} \right\rvert}
+
\frac{ \dfrac{\Sigma_{in}(E)}{\Sigma_t (E)} Q(E) } {\Sigma_t (E)} ,
\label{eq52b}
\end{equation}
%\end{strip}

\noindent where average-logarithmic energy decrement of inelastic scattering $\xi_{in}$ is introduced by analogy to standard neutron moderation theory (see e.g.~\cite{Feinberg1978, BartolomeyBat1989, Shirokov1998, Rusov2017}), but through inelastic neutron scattering law (\ref{eq50}). After performing calculations similar to calculations performed for $\xi_{el}$ in (\ref{eq25}), we obtain the following expression:

\begin{equation}
\xi_{in} = 
\frac{\widetilde{\alpha}_1}{1-\widetilde{\alpha}_1}
\mathrm{ln}\left(\widetilde{\alpha}_2^2\widetilde{\alpha}_1\right)+
\frac{\left(1-\widetilde{\alpha}_2^2\widetilde{\alpha}_1\right)}
{\widetilde{\alpha}_2^2(1-\widetilde{\alpha}_1)}
+
\mathrm{ln} \left(
\frac{E}
{E+\frac{1}{A}\frac{3}{2}kT}
\right) .
\label{eq54new}
\end{equation}

Therefore, considering elastic and inelastic neutron scattering it is possible to obtain an expression for total moderating neutrons flux density:
\begin{equation}
\Phi(E) = \Phi_{el}(E) +\Phi_{in}(E)
.
\label{eq53}
\end{equation}

And the neutron spectrum will be defined by a standard expression:
\begin{equation}
\rho(E) = \frac {n(E)}{\int\limits_0^\infty {n(E')\mathrm{d}E'}} =
\frac{\frac{1}{\sqrt{2E} }\Phi(E)} 
{\int\limits_0^\infty { \frac{1}{\sqrt{2E'}} \Phi(E') \mathrm{d}E' }}
.
\label{eq54}
\end{equation}

Let us note, that considering discreteness of energy levels of moderating medium nuclei at which neutrons are scattered inelastically, it is obvious that the inelastic scattering is possible not at any value of neutron kinetic energy, but with its kinetic energy above a threshold given by the following expression (see (\ref{eq27})):
\begin{equation}
E_{in}^{thold} = \frac{A+1}{A}E_{min}^*
,
\label{eq55}
\end{equation}
where $E_{min}^*$ is the minimal energy of excitation of the scattering nucleus.

This is confirmed by experimental results, e.g. according to~\cite{Vlasov1971}, neutron inelastic scattering at heavy nuclei is observed only at neutron energy higher than several hundreds kiloelectronvolts. And for light nuclei -- at energies higher than one or several megaelectronvolts.

For fission reactor environments $Q(E)$ is determined by fission spectrum of the fission nuclide or fission nuclides combination, which according to~\cite{Weinberg1958, Stacey2001, Fedorov1961, Shirokov1972, Rusov2017} may be given by the following expression:

\begin{equation}
Q(E) = \widetilde{Q} \cdot q \cdot \exp{(-aE)} \cdot \mathrm{sh} (\sqrt{bE})
,
\label{eq56}
\end{equation}
where $a$, $b$ and $q$ -- constants, given in table~\ref{table1} below, $E$ -- neutron energy nondimensionalized by 1 MeV, $\widetilde{Q} = \int\limits_0^\infty {Q(E)\mathrm{d}E}$ -- total neutrons quantity generated in a unit volume in a unit time.

\begin{table}[htb!]
\centering
\caption{Constants determining the fission spectrum for main reactor fission nuclides~\cite{Fedorov1961}.}
\label{table1}
\renewcommand{\arraystretch}{1.7}
\begin{tabular}{| c | c | c | c | c |}
\hline
~ & \textbf{U${}^{235}$} & \textbf{Pu${}^{239}$}  &
\textbf{U${}^{233}$} & \textbf{Pu${}^{241}$}
\\  \hline
$a$ & 1.036 & 1 & 1.05 $\pm$ 0.03 & 1.0 $\pm$ 0.05
\\  \hline
$b$ & 2.29 & 2 & 2.3 $\pm$ 0.10 & 2.2 $\pm$ 0.05
\\  \hline
$q$ & 0.4527 & $\sqrt{\frac{2}{\pi e}} = 0.48394$ & 0.46534 & 0.43892
\\  \hline
\end{tabular}
\end{table}

%------------------------------------------------------------------------------------------------------------------------------------
%------------------------------------------------------------------------------------------------------------------------------------
%------------------------------------------------------------------------------------------------------------------------------------

\begin{figure}[htb!]
\centering
\includegraphics[width=8cm]{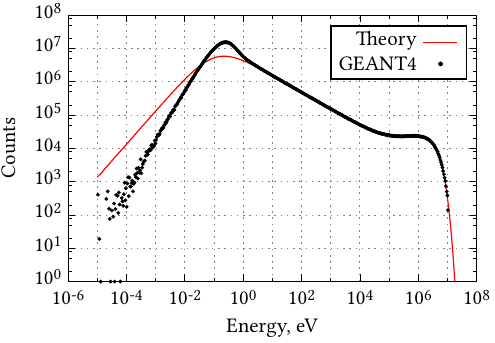}
\caption{Comparison of the theoretical and Monte-Carlo spectra of moderating
neutrons in hydrogen at 1200~K.}
\label{fig-moderation-H1}
\end{figure}

Fig.~\ref{fig-moderation-H1} shows the energy spectrum of neutrons moderating
in hydrogen medium calculated using the Eq.~(\ref{eq51}). The neutron source
was determined by the Eq.~(\ref{eq56}) for uranium-235 (Table~\ref{table1}),
the medium temperature was set 1200~K. The cross-sections of neutron reactions
in hydrogen were taken from the ENDF/B-VIII.0 database~\mbox{\cite{ENDFBVIII0}}.
This spectrum is compared to the one obtained by Monte-Carlo simulation using
GEANT4 package~\mbox{\cite{GEANT42003}}, and demonstrates a rather good agreement.

The comparison of the two neutron spectra in hydrogen moderator, obtained using
different methods and presented in figure \mbox{\ref{fig-moderation-H1}}
demonstrates a good agreement between them.

\begin{figure*}[tb!]
\centering
\includegraphics[scale=1]{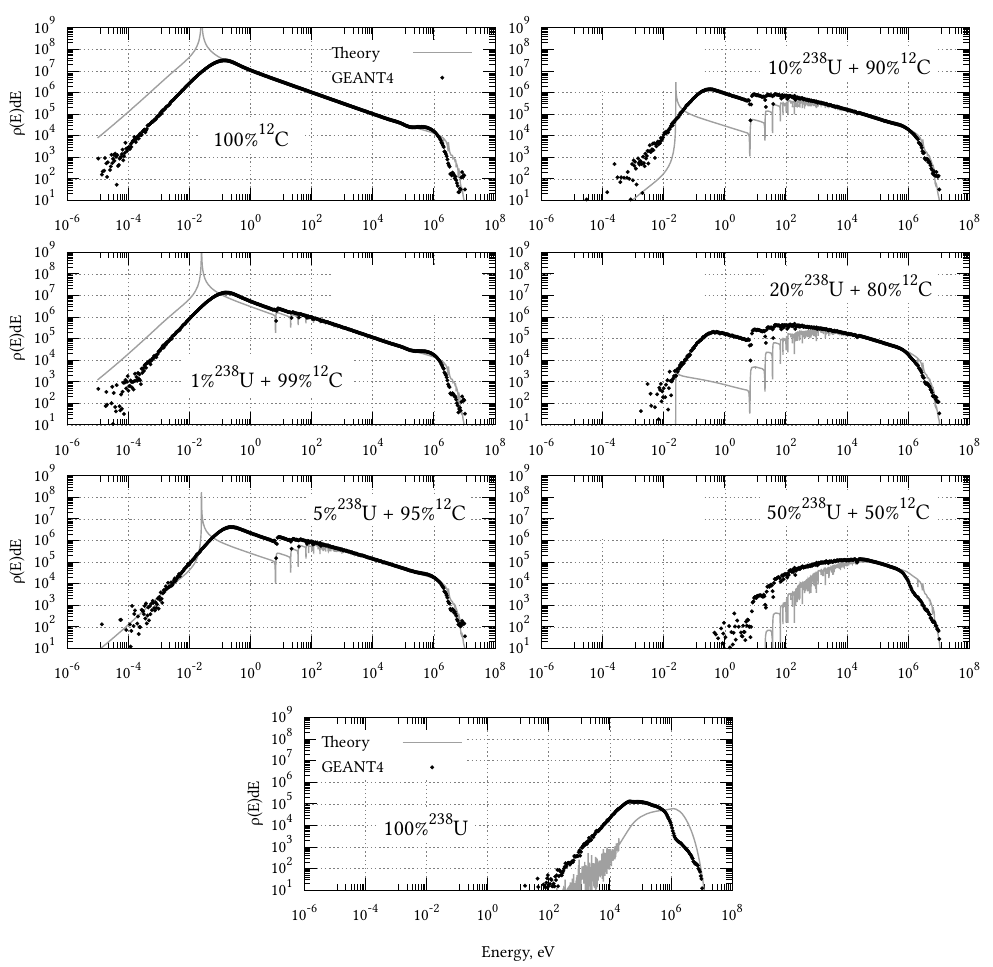}
\caption{Energy spectra of the moderating neutrons in uranium-carbon media
of different composition calculated using Eq.~(\ref{eq54}) (Theory, solid line)
and using GEANT4 software (black dots). The initial spectrum was that of
fission (\ref{eq56}). The medium temperature is 600~K.}
\label{fig-GEANT4-U238}
\end{figure*}

As it was noted in our recent paper~\cite{Rusov2017}, the analytic expression
obtained within the new theory of neutron moderation, describes the entire
spectrum of moderating neutrons -- i.e. in the entire band of possible
energies depending on the moderator composition and temperature. It may contain
two maxima (high-energy and low-energy) or a single maximum (high-energy for
fast neutrons or low-energy for thermal neutrons, or an intermediate one). Such
transformation of the theoretical spectrum is demonstrated below by the series
of images depicting the spectra of neutrons moderating in uranium-carbon media
of different composition.
	
\begin{figure}[tb!]
\centering
\includegraphics[width=8cm]{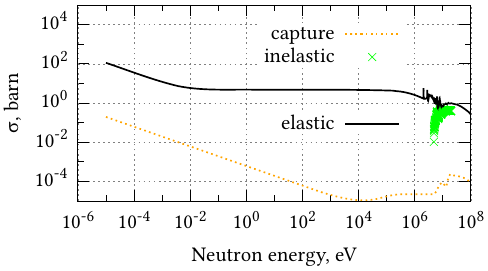}
\caption{The microscopic cross-sections for the neutron reactions in carbon at 600~K.}
\label{fig-XS-C12}
\end{figure}

\begin{figure}[tb!]
\centering
\includegraphics[width=8cm]{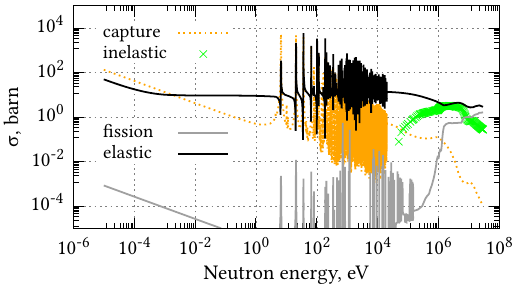}
\caption{The microscopic cross-sections for the neutron reactions in uranium-238 at 600~K.}
\label{fig-XS-U238}
\end{figure}

\mbox{Fig.~\ref{fig-GEANT4-U238}} shows the energy
spectra of neutrons moderating in homogeneous uranium-carbon medium. The 
theoretical curves were calculated using Eq.~(\ref{eq54}), and the 
Monte-Carlo simulated ones were obtained using GEANT4 software. The neutron
source was determined by Eq.~(\ref{eq56}) for uranium-235 in all cases.
The neutron reactions cross-sections for uranium-238 (Fig.~\ref{fig-XS-U238})
and carbon (Fig.~\ref{fig-XS-C12}) were taken from 
ENDF/B-VIII.0~\cite{ENDFBVIII0}.

From comparison of the corresponding curves in \mbox{Fig.~\ref{fig-GEANT4-U238}}
it may be seen that the theoretically calculated spectra are in good 
qualitative agreement with
the Monte-Carlo simulated ones. Both curves match in a wide range for a
100\% carbon medium. In the low energy range the curves somewhat differ, but
the slope in log-log scale remains the same. With increase of the $^{238}U$
portion in the moderator, our theoretical model gives noticeably lower values
below few keVs, which is a subject of further investigation. The high energy
part of the spectrum as well as the slopes in log-log scale are still very
close though.

The series of images demonstrates the gradual transformation of the neutron
spectrum from fast to thermal with increase of the carbon percentage in the
moderating mixture.

Let us mention the sharp peak in the thermal part of the theoretical spectrum
(seen in \mbox{Fig.~\ref{fig-GEANT4-U238}}). It is
associated with the logarithmic energy decrement $\xi$ crossing zero, since the
neutron flux density formally tends to infinity in this case (see (\ref{eq51})
or (\ref{eq52b})). As it was noted in our recent paper~\cite{Rusov2017}, for the new
scattering law, the energy decrement is not constant, in contrast to the
standard moderation theory. Instead, it depends on the neutron energy and
moderator temperature (see e.g. (\ref{eq54new})). In the energy range below
the thermal energies the logarithmic energy decrement $\xi$ becomes negative.
This seems reasonable, since such neutrons would obviously \textit{gain} some
extra energy when interacting with the thermally moving nuclei. So the neutrons
are "pushed" away from the zero energy, and our new theory of neutron
moderation "captures" this feature.

Let us consider the point which corresponds to the zero value of the
logarithmic energy decrement. The theoretical expression for the neutron
flux is divergent at this point. According to the definition of the
logarithmic energy decrement (e.g. \cite{Feinberg1978,BartolomeyBat1989,Shirokov1998,Rusov2017}), the energy
of such neutrons after the interaction remains the same as before. So the
neutrons with such energy tend not to change it in the course of interaction
with moderator nuclei. Therefore such neutrons would accumulate.

However if we look at the kinematics of two-particle elastic scattering, the
case when the light incident particle after collision has the same energy as
before, corresponds to the only point of the momentum diagram -- when both
particles after collision move along the same line in the L-system. I.e. 
$P_1 ^{(L)} \parallel P_2 ^{(L)}$, where $P_1 ^{(L)}$ and $P_2 ^{(L)}$ are
the momenta of the incident and scattering particles in L-system after
collision (e.g. \cite{Sitenko1993}). This point corresponds to the solution $P_1 ^{(L)} = P_{10}^{(L)}$
and $P_2 ^{(L)} = 0$, $P_{10}^{(L)}$ is the incident particle momentum in 
L-system after collision. This solution is traditionally considered as valid,
but it is obviously not physical, since the situation when the classical
particle is scattered on another one, while preserving its momentum vector,
is not possible. So the point at which the logarithmic energy decrement
is zero, should be excluded from consideration as non-physical.

It is interesting though that when considering the inelastic scattering
through the stage of compound nucleus, this point cannot be excluded.
The neutron can be ejected from the compound nucleus with the same
momentum vector it had before. This problem is smoothed over by the fact that
the inelastic scattering is a threshold reaction (with the threshold about
$10^{5}$~keV), and its cross-section is zero for the neutron energy
corresponding to the point of $\xi = 0$.

\section{Conclusions}

For the first time a general analytic expression for inelastic neutron scattering law considering temperature of the moderating medium as a parameter is obtained for isotropic neutron source. Also analytic expressions for neutron flux density and neutron moderation spectrum are obtained for isotropic neutron source in neutron moderating and absorbing media (e.g. different reactor media), containing nuclei of different kinds and also depending on medium temperature.

The obtained expressions for moderating neutrons spectra open a new way to
interpret the physical nature of processes that determine the shape of
neutron spectrum in low-energy range. We also found how the scattering
cross-sections and the logarithmic energy decrement behavior impact the
formation of the low-energy maximum in the moderating neutron spectrum.
The nature of this maximum is coupled with non-stationary neutron system moderation by scattering at thermalized moderation medium nuclei system. Therefore its peculiarities cannot be explained within classic approach considering neutrons system in thermodynamic equilibrium and obeying Maxwell's distribution.

It is noteworthy the presented theoretical model allows to obtain the neutron
moderation spectra for the case of non-equilibrium moderator, i.e. without
introducing the moderator temperature into the model. This is also true for the
basic theoretical model~\cite{Rusov2017}. Indeed, our model remains rather generic
until the Eq.~(\ref{eq46}), where an assumption about the thermal equilibrium
of the moderator nuclei is made, and the Maxwell distribution is applied 
(Eq.~(\ref{eq47})). Then, by averaging of the expression (\ref{eq46}) across
the Maxwell's distribution, the moderator temperature is introduced. However,
such assumption is not so indispensable. Suppose, the kinetic energy
distribution of the moderator nuclei $P_M \left( E_N^{(L)} \right) dE_N^{(L)}$ is
known. Then the averaging of the expression (\ref{eq46}) over this distribution
would only change the $\frac{3}{2}kT$ in (\ref{eq48}) to $\overline{E}_N^{(L)}$, 
where $\overline{E}_N^{(L)} = \int \limits_0^{\infty} E_N^{(L)} P_M \left( E_N^{(L)} \right) dE_N^{(L)}$
is the mean kinetic energy of the moderator nuclei. This way, substituting
$\frac{3}{2}kT$ by $\overline{E}_N^{(L)}$ in all subsequent equations, one
would obtain a more general form of neutron moderation theory.

The energy spectra of neutrons moderating in homogeneous hydrogen and
uranium-carbon media are presented. The graphs were obtained using the
analytical expression~(\ref{eq54}) and the GEANT4 Monte-Carlo code~\cite{GEANT42003}.
The comparative analysis of the neutron spectra obtained using both methods
reveals a good agreement.

Finally, a significantly different behavior of elastic and inelastic scattering
cross-sections in different moderators (see e.g. ENDF/B-VIII.0 or~\mbox{\cite{Stacey2001}})
provides opportunities for experimental investigation of their influence on the
shape of maximum in neutron moderation spectrum in low-energy range, and also
for the experimental testing of the obtained analytic expressions.

\bibliographystyle{unsrtnat}
\bibliography{Tarasov-ModerationInelastic}

\end{document}